\documentclass[twocolumn,floatfix]{revtex4}
\usepackage{graphicx}
\usepackage{dcolumn}
\usepackage{amsmath}

\usepackage{amsmath,amsfonts,amssymb}
\usepackage{wrapfig}
\usepackage{graphicx}
\usepackage{bbm}
\usepackage{graphics}

\def \beq {\begin{equation}}
\def \eeq {\end{equation}}
\def \ba {\begin{eqnarray}}
\def \ea {\end{eqnarray}}

\begin{document}

\hsize\textwidth\columnwidth\hsize\csname@twocolumnfalse
\endcsname

\title{Cryogenic High-Frequency Readout and Control Platform for Spin Qubits}
\author{J. I. Colless and D. J. Reilly$^*$}
\affiliation{ARC Centre of Excellence for Engineered Quantum Systems, School of Physics, The University of Sydney, Sydney, NSW 2006, Australia}

\begin{abstract}
We have developed a cryogenic platform for the control and readout of spin qubits that comprises a high density of dc and radio frequency sample interconnects based on a set of coupled printed circuit boards. The modular setup incorporates 24 filtered dc lines, 14 control and readout lines with bandwidth from dc to above 6 GHz, and 2 microwave connections for excitation to 40 GHz. We report the performance of this platform, including signal integrity and crosstalk measurements and discuss design criteria for constructing sample interconnect technology needed for multi-qubit devices.  
\end{abstract}
\maketitle
\section{Introduction}
Nanoscale circuits that enable coherent manipulation and readout of single electron spin-states are of interest as platforms for constructing quantum information technology \cite{Loss:1998via,Kane:1998wha,Hanson:2007eg}. These qubit devices are operated at cryogenic temperatures by controlling electron energy levels using nanosecond voltage pulses applied to metal electrodes on the surface of a semiconductor heterostructure \cite{Petta:2005kn,Koppens:2006kz}. At present an evolution is underway, from single-qubit architectures that have demonstrated state preparation, arbitrary superposition \cite{Petta:2005kn,Koppens:2006kz,Nowack:2007du,Foletti:2009hka}, and single-shot readout \cite{Elzerman:803014,Amasha:2008ky,Barthel:2009hx,Morello:2010ga}, to {\it multi-}qubit devices needed to quantify entanglement and perform computation via the parallel operation of several quantum gates \cite{vanWeperen:2011fl, Brunner:2011gma,Nowack:2011}. Scaling from single to few qubits, in addition to the major scientific challenges, also requires technical advances such as the  development of new hardware and methods for enhancing readout, control, and noise mitigation in multi-qubit cryogenic setups. 

Crosstalk between control signals presents a challenge for scale-up of spin qubit devices, increasing error rates for single qubits and opening new channels for decoherence in multi-qubit architectures. In particular, the broadband nature of control waveforms, which are typically large-amplitude rectangular `dc'  pulses with sub-nanosecond rise-times, resemble a mixed-signal environment in which digital logic circuits can interfere with sensitive analog systems \cite{Su:1993vm}. Maintaining a high degree of readout and control signal fidelity under these conditions is necessary if spin-qubit architectures are to reach the low hardware error thresholds required for quantum error correction \cite{Preskill:1998vu}.

Many of these technical challenges are not unique to quantum devices and are common place in the context of commercial monolithic microwave integrated circuit (MMIC) implementation and packaging. In contrast however, interconnect solutions for spin-qubit device development require cryogenic and high magnetic field operation together with a flexibility that allows for the many iterations of a design, fabrication, and measurement cycle. For instance, interconnects are required to accommodate the regular changing of sample chips of different size and bonding configuration. 
\begin{figure*}
\includegraphics[scale=0.30]{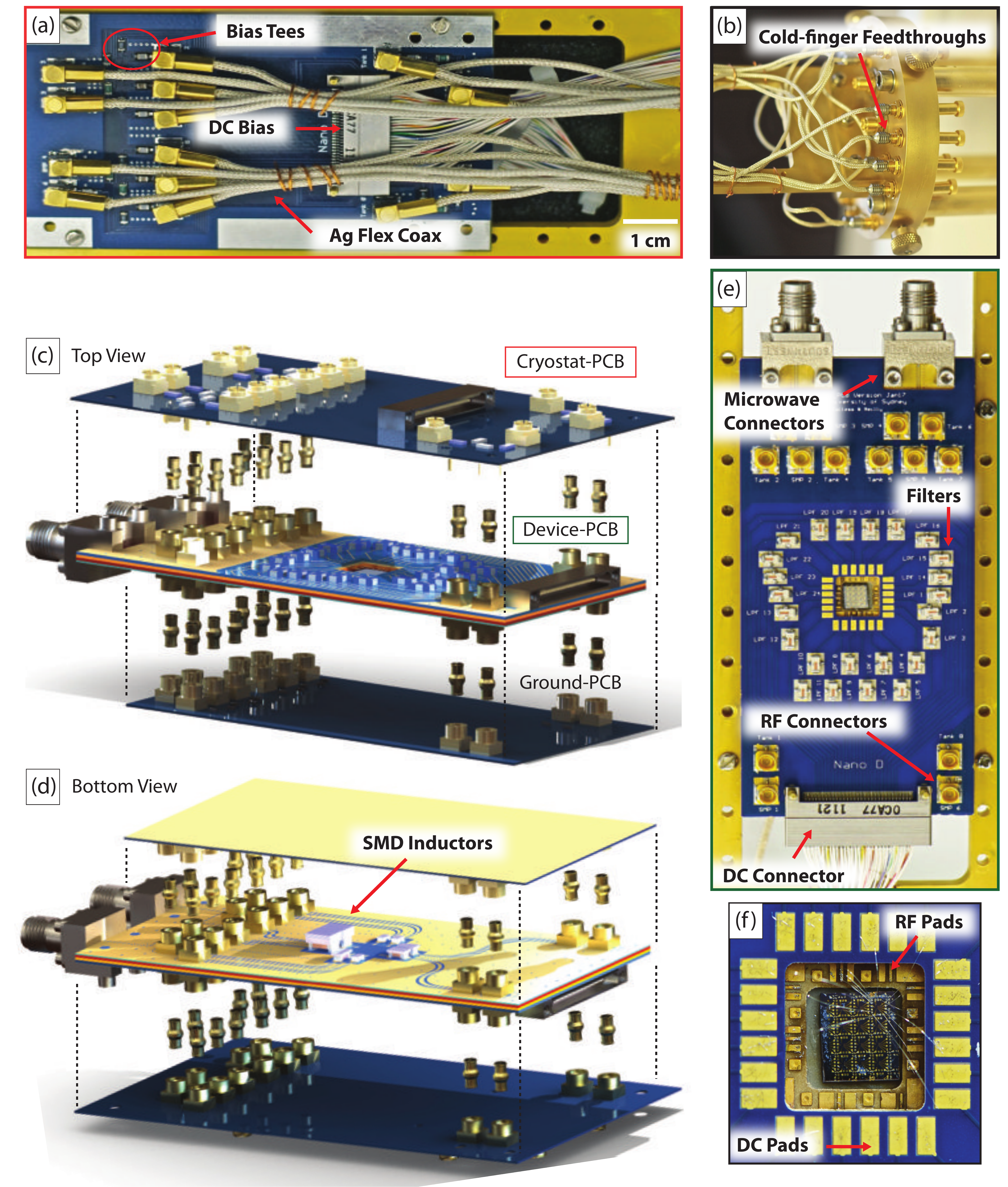}
\caption{\label{fig:photos} Modular PCB architecture incorporating a high density of interconnects needed for multi-qubit readout and control.  \textbf{(a)} The cryostat-PCB is fixed to the cold finger at the mixing chamber of a dilution refrigerator. \textbf{(b)} Shows the high-frequency custom cables and feed-throughs entering the cold finger.
\textbf{(c)} Top view and \textbf{(d)} bottom view cartoon of the coupled PCB set (rendered using Solidworks CAD software package). \textbf{(e)} The device-PCB, which houses the  chip, connects to the cryostat-PCB with the use of MSMP connectors and bullets.  The device-PCB has 31 filtered dc lines, matching circuits, and 2 microwave frequency connections together with 14 rf lines that are passed through from the cryostat-PCB. The ground-PCB  pushes on to the device-PCB from the other side, allowing for  `make-before-break' connections. \textbf{(f)} Terraced bond pads and recess for housing the qubit chip. }
\end{figure*}
Quantum coherent circuits are also different to typical MMIC architectures in that they can be sensitive to very broadband noise and interference (hertz to terahertz) which increases the device electron temperature and, when strong enough, can artificially drive transitions between qubit energy levels \cite{Gustavsson:2008gb}, lead to photon assisted tunnelling \cite{Kouwenhoven:1994ww}, or create bias currents from rectification \cite{Switkes:1999uva}. For spin qubits, even small amplitude noise or crosstalk (of the order of nanovolts) reduces the fidelity of quantum gate operations by introducing uncontrolled fluctuations of the electron potential defined electrostatically using metallic surface electrodes. Error suppression methods that dynamically decouple environmental fluctuations can serve to mitigate noise on control lines, but these introduce an additional computational overhead and fail in the limit of white noise derived from thermal sources  \cite{Khodjasteh:2007bu}.

Here we report a low-noise readout and control platform that incorporates the high density of interconnects needed to operate multi-qubit devices at cryogenic temperatures. The modular platform makes use of a series of microwave printed circuit boards (PCBs) that connect together to enable ease of sample exchange. The main device-PCB is a 5-layer laminate that electrically partitions dc, radio-frequency (rf), and microwave signals using ground planes and a dense array of vias. Such partitioning is shown to strongly suppress high-frequency crosstalk in device architectures that require a high-density of signal interconnects. Characterizing our setup, we present signal fidelity measurements at cryogenic temperatures and compare these to EM circuit models and simulations. Although developed specifically for spin qubits, we anticipate that the results reported here are of general interest for experiments that involve high-frequency measurements of nanoscale devices at cryogenic temperatures.    
\section{Coupled Printed Circuit Boards}
We first describe the setup of the circuit boards and their interconnects, including details of the cryostat wiring and filters used to suppress noise in our system. The 3 PCBs comprise a cryostat-PCB, in thermal contact with the mixing chamber of a dilution refrigerator, a 5-layer device-PCB that houses the wire bonded sample chip, and a ground-PCB that allows `make-before-break' connections of all high-frequency lines to protect the device from electrostatic discharge.
\subsection{Cryostat-PCB and Wiring}
Fast voltage pulses for spin qubit control are produced using room temperature waveform generators and transmitted to the sample chip using semi-rigid coaxial cables \cite{coax} thermally anchored in the dilution refrigerator using attenuators. The chip is mounted inside a light-tight cold-finger attached to the mixing chamber stage of a `cryo-free' dilution refrigerator with base temperature $\sim$ 18 mK \cite{Leiden}. Impedance-matched microwave filters ($Z_0$ = 50 $\Omega$), constructed using magnetically lossy epoxy \cite{Eccosorb}, further thermalize and limit  the frequency bandwidth of coaxial connections. Multi-stage cryogenic $RC$ filters are used on all low-frequency wiring, which are shielded between room temperature and the cold finger.  Using Couloumb blockade thermometry we measure an electron temperature below 40 mK with this high-frequency setup.

Custom cryogenic cables connect MCX- and SMA-type coaxial connectors at the cold-finger feed-throughs [see Fig. 1(b)] to the cryostat-PCB shown in Fig. 1(a). These custom cables are non-magnetic, hand-formable, and consist of a semi-rigid copper inner conductor, followed by a teflon dielectric wrapped with a silvered copper foil and a silver braid that serves as  the outer conductor. High-frequency connectors on the cryostat-PCB are MMCX-type. 

The cryostat-PCB is in strong thermal contact with the gold-plated, high-purity copper cold finger and remains attached to the refrigerator. On-board bias tees constructed from surface mount resistors and capacitors add true-dc, or low-frequency signals to the high-frequency readout and control lines. A `nano-D' connector \cite{nano} [shown in Fig. 1(a)] is used to connect these dc lines to the bias tees on the cryostat-PCB.

The set of interconnecting PCBs is shown as a rendered cartoon in Fig. 1(c) and Fig. 1(d). The boards are connected to each other using mini-SMP connectors \cite{smp} and interconnect coaxial `bullets' that allow a radial misalignment of  $\sim$ 0.4 mm and an axial misalignment of $\sim$ 0.7 mm. The grounding-PCB is mated with the device-PCB during sample bonding and transport and ties all high-frequency connections to a common ground via 500 k$\Omega$ resistors that dissipate high voltage spikes that can otherwise shock the device. To make connection with the cryostat wiring the device-PCB is first mated with the cryostat-PCB before removal of the ground-PCB. The force required to separate the bullets from their connectors is specified such that all bullets remain attached to the cryostat- and grounding-PCBs, rather than the device-PCB. In this way the device is connected to a high resistance ground throughout.
\begin{figure*}
\includegraphics[scale=0.3]{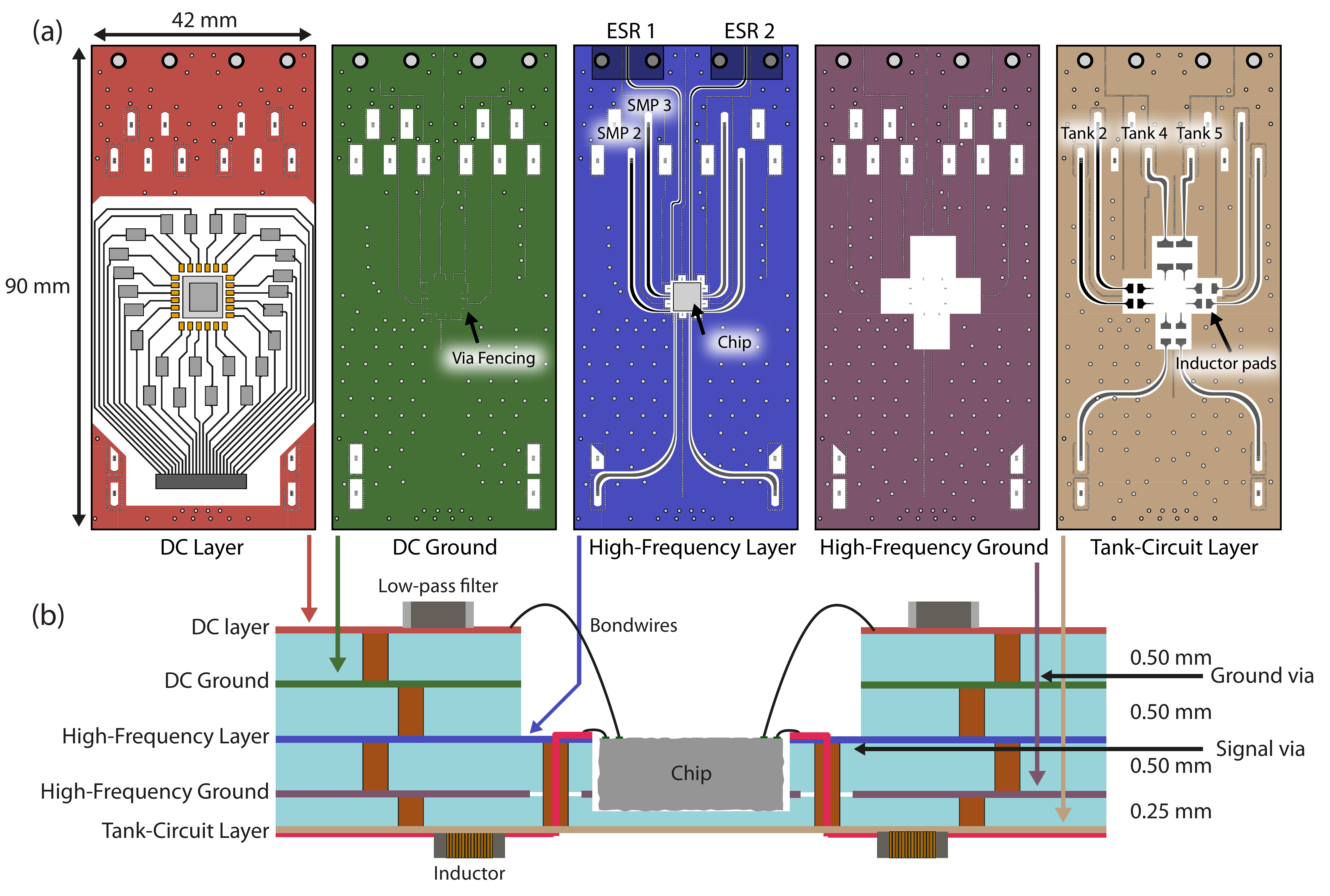}
\caption{\label{fig:test}  \textbf{(a)} Layout of the 5 individual device-PCB layers showing ground-planes that are used to electrically partition high-frequency connections.  \textbf{(b)} Cross-section of the device-PCB in a region close to the chip. The square recess in the device-PCB is created by layers of laminate shaped to frame the chip and allow for short bond wires for high-frequency interconnects. The low dielectric constant of Rogers 3003 laminate suppresses crosstalk between adjacent signal tracks (see text for details).}
\end{figure*}
\subsection{Device-PCB}
The chip is mounted in a square recess created in the circuit board by the use of multiple framing layers of Rogers 3003 laminate bonded together to build up the device-PCB, shown as a photograph in Fig. 1(e) and schematically in Fig. 2.  Bond pads for the high-frequency signals emerge close to the chip on the high-frequency layer of the device-PCB, with dc bond pads located further away, on the top dc-layer. This creates a terraced bond pad structure that reduces the bondwire length for high-frequency connections [see Fig. 1(f)]. The Rogers laminate has a thermal expansion coefficient matched to copper and exhibits a small variation in dielectric constant with temperature (+13 ppm / degree), ensuring that the impedance of planar transmission lines does not change when cooling. The laminate also exhibits a low loss (0.0013 dissipation factor) and relatively high thermal conductivity, making it well suited to microwave cryogenic applications. We note that we have performed many thermal cycles of this PCB without degradation.

Metallic features on the device-PCB are defined using electroless-nickel electroless-palladium immersion gold (ENEPIG) finish which ensures strong wire bond adhesion. The board remains essentially non-magnetic as only a trace amount nickel is used as a sticking layer during PCB metal deposition and all components are non-magnetic. Connecting the ground planes and signal layers are a large number of `plated through' vias that are plugged with epoxy so as to not trap gas that may otherwise act as a virtual leak under vacuum. These vias also suppress any parallel-plate capacitor resonance modes produced by metallic layers in the PCB  \cite{Tischler:2003tp,Yuasa:2004uf}.

\begin{figure*}
\includegraphics[scale=0.3]{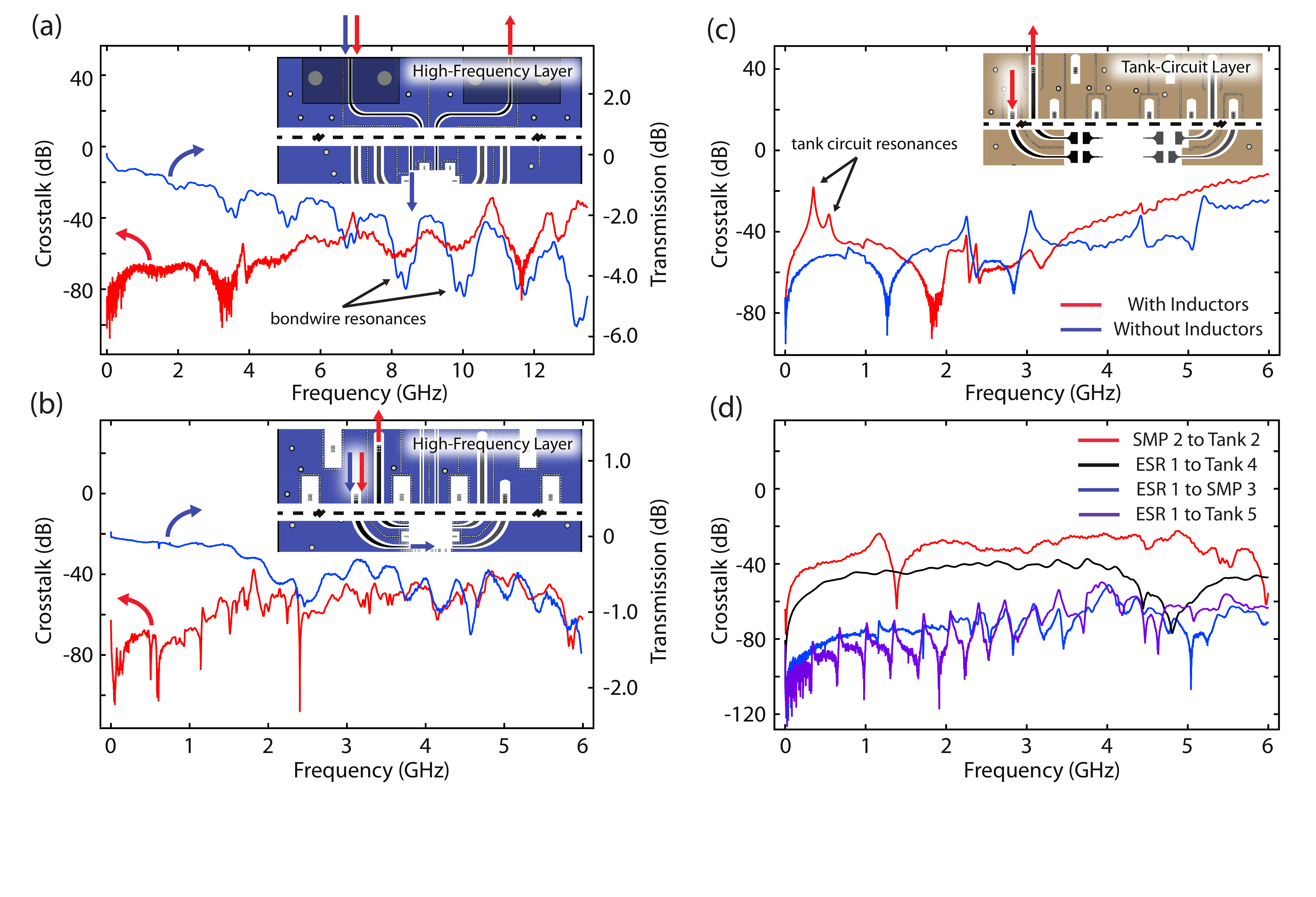}
\caption{\label{fig:crosstalk} \textbf{(a)} Crosstalk between the two ESR signal tracks on the high-frequency layer of the device-PCB (image of the PCB layer shown as an inset). The crosstalk between the signal lines (shown in red) remains below -40 dB for frequencies below 10 GHz.  Transmission data for the ESR microwave lines (shown in blue) is obtained by linking the lines together using aluminium bond wires. These are required for the transmission measurement but introduce small parasitic resonances. The data shown is the total loss, from ESR 1 to ESR 2, divided by 2. \textbf{(b)} Crosstalk between the nearest neighbour high-frequency lines (with ports indicated on the  image shown as an inset). Transmission is again measured by linking the end of the tracks with bond wires.  \textbf{(c)} Crosstalk between two tank-circuit transmission lines (ports again shown in the inset). Crosstalk measured without inductors soldered to the pads is shown in  blue. Measurements taken with inductors in place are shown in red. We note corresponding peaks in the crosstalk spectrum at the resonant frequencies of the individual tank circuits. \textbf{(d)} Measured crosstalk between various different transmission lines on the device-PCB [see Fig. 2(a) for port labels]. Maximum crosstalk occurs between vertically adjacent tank-circuit lines and high frequency tracks (shown in red). The majority of crosstalk is produced in the region close to the chip where the ground-planes have been removed.}
\end{figure*}

The low dielectric constant of Rogers 3003 ($\epsilon_r$ =  3) allows the design to minimize crosstalk despite the high density of planar transmission lines. Constrained by the minimum feature size compatible with PCB manufacture, a low dielectric constant allows an impedance of $Z_0 \sim$ 50 $\Omega$ to be maintained by having the ground-plane positioned a close distance underneath the signal tracks relative to the distance between neighbouring tracks. In this way, electric field lines stemming from the signal lines mostly terminate on the ground-plane beneath the transmission lines rather than terminating on adjacent signal tracks which would otherwise increase coupling \cite{Imai:2009tn}.

The layout of the individual layers of the device-PCB are shown in Fig. 2(a). The dc layer (top surface) of the device-PCB has 24 low-frequency signal tracks that connect bond pads [see Fig. 1(f)] to a 31-pin `nano-D' dc-connector, shown in Fig. 1(e). The remaining 7 wires on the connector provide additional connections for thermometry, active device power, or cold light-emitting diodes. Each low frequency line is first filtered at the mixing chamber using low-pass $RC$ stages embedded in magnetically lossy epoxy \cite{Eccosorb} and shielded before entering the light-tight cold finger housing. This combination provides more than -60 dB of noise suppression for frequencies from 700 Hz to above 50 GHz. We additionally make use  of 7-stage surface mount low-pass filters (80 MHz cutoff frequency \cite{mini}) located on the PCB close to the device, to suppress high-frequency crosstalk from rf  to the dc lines [see Fig. 1(e)]. We note that mounting chip capacitors in place of these filters (on the PCB) can introduce parasitic resonances in the frequency band of control signals. 

High-frequency signals are fed to the chip via 14 separate coplanar waveguides embedded on distinct layers of the device-PCB. Contact to the high-frequency layer is made using mini-SMP connectors mounted on the top (and bottom) surface of the device-PCB. The central pin of these connectors is soldered to a via that makes contact to either the high-frequency layer or tank-circuit layer of the device-PCB [see Fig. 2(a)]. These ground-covered coplanar waveguides have low dispersion and are adiabatically tapered from the $Z_0$ = 50 $\Omega$ SMP connectors to $\sim$ 83 $\Omega$ at the bond pads to minimise impedance mismatch with the bonding wire geometry. Ground-planes separate each signal layer together with a fencing-via technique \cite{Ponchak:1998tm,Wu:2003wa}, that effectively terminates electric field lines from high-frequency signal tracks in order to suppress crosstalk (see discussion below).  

Electron spin resonance (ESR) is a method needed for spin qubit manipulation and typically requires microwave frequencies for excitation. To enable ESR we make use of 2 edge-mounted 2.40 mm microwave launchers \cite{esr}. These connectors maintain good impedance matching and signal integrity to 40 GHz and are connected to the chip using coplanar waveguide structures (ESR 1 and ESR 2) on the high-frequency layer of the device-PCB [see Fig. 2(a)]. These waveguides are again well isolated from other signal lines using fencing-vias.  

Finally, on the bottom side of the device-PCB is the tank-circuit layer which contains solder pads for incorporating surface mount components in series with coplanar waveguides. We typically mount chip inductors here to implement $LC$ tank-circuits for the purpose of impedance matching to charge sensors needed in rf reflectometry \cite{Reilly:2007ig} for spin readout \cite{Barthel:2009hx}. The 8 solder pads differ in size to accommodate the range of surface mount gauges. The presence of metal structures proximal to the inductor mounts are minimized to reduce parasitic capacitance (measured to be $\sim$ 0.2 - 0.3 pF). 
 \section{Crosstalk and Signal Fidelity Measurements}
Having described the layout of the cryostat- and device-PCB we now present low temperature measurements characterizing the crosstalk and transmitted signal fidelity of the coupled PCB architecture. Measurements are made with a calibrated vector network analyzer \cite{PNA} at a temperature $T \sim$ 5 K using a high-frequency cryogenic probe-station \cite{Lakeshore}. We have verified that microwave $S$-parameters do not change when the PCB is cooled further to milli-Kelvin temperatures. 
 
Beginning with the edge-mounted microwave launchers used for ESR, Fig. 3(a) shows the crosstalk (red) and transmission performance (blue) of the device-PCB. Crosstalk from the two planar transmission lines is determined via a measurement of $S_{21}$ between the unconnected ports, ESR 1 and ESR 2. We find a maximum crosstalk of $\sim$ -40 dB at frequencies above 10 GHz. The transmission performance [shown in Fig. 3(a)] is determined by connecting the bond pads with long bond wires where they terminate close to the chip cavity and again measuring $S_{21}$ between ports ESR 1 and ESR 2. The bond-wires are required to perform a transmission measurement but lead to addition loss and parasitic resonances from the bond wire partial inductance and stray capacitance.  Without losses from the bond-wires, numerical simulations \cite{HFSSQ3D} indicate that transmission drops to $S_{21} \sim$ -3 dB at $\sim$ 13 GHz. We note that these coplanar waveguides have the shortest distance between them of all the signal lines on the PCB and exhibit the strongest crosstalk.  
\begin{figure*}
\includegraphics[scale=0.3]{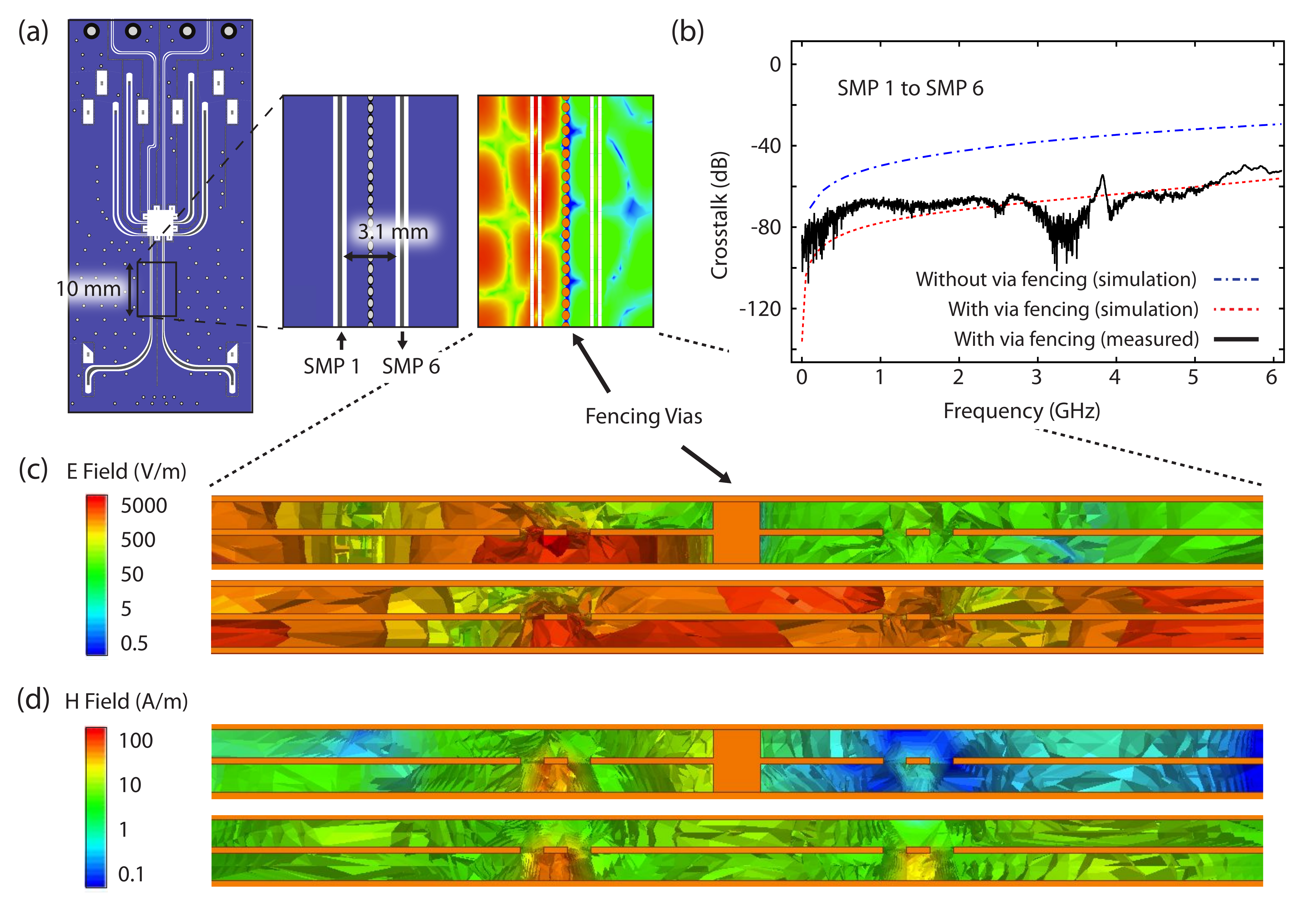}
\caption{\label{fig:sims}We examine the effect of the fencing-via technique on the EM coupling and crosstalk between signal lines.  \textbf{(a)} Layout of the high-frequency layer of the device-PCB showing zoom and electric field strength obtained using EM simulation software \cite{HFSSQ3D} [see scale bar in (c)]. An input voltage amplitude of 1 V is applied to the port SMP 1 at a frequency of 40 GHz.  \textbf{(b)} Crosstalk between high-frequency ports SMP 1 and SMP 6 in a bandwidth 0 - 6 GHz. Simulation software is used to evaluate crosstalk when fencing-vias are used (red dashed line) and not used (blue dashed line). \textbf{(c)} Numerical simulation of the field strength in a cross-section of the high-frequency layer of the device-PCB, again at 40 GHz.  The magnitude of the electric field is shown with (top) and without (bottom) fencing-vias.  \textbf{(d)} Shows magnetic field strength comparing the coupling between ports, with and without fencing-vias for conditions as in (c).  A reduction in the coupling strength of electric and magnetic field is seen when fencing-vias are implemented. A via diameter of 0.38 mm is used with a via centre-to-centre spacing of $\sim$ 0.6 mm.}
\end{figure*}
The 14 transmission lines that use mini-SMP connectors to make contact with the high-frequency layer are designed for carrying control pulses with a bandwidth below $\sim$ 6 GHz. The performance of these interconnects is shown in Fig. 3(b) for two neighbouring lines. A loss in transmission of $S_{21}\sim$ -1 dB is observed, again using bond-wires to connect signal tracks near the chip cavity to allow the transmission measurement. Crosstalk between adjacent lines remains below -40 dB for frequencies up to 6 GHz.

Readout interconnects on the tank-circuit layer are investigated via measurements with, and without, surface mount inductors soldered to nearest neighbour transmission line pads [see Fig. 3(c)]. In both cases, readout line crosstalk is larger relative to the performance of the lines on the high-frequency layer shown in Fig. 3(b). This is because, in the case of the tank-circuit layer, the signal tracks are on the back surface of the PCB and do not have a covering ground plane to enhance shielding. Tank-circuit resonators used for readout typically operate however, at frequencies below 3 GHz, where crosstalk remains less than -40 dB. The use of inductors to create tank-circuit resonators leads to an increase in crosstalk at the resonant frequency of the respective tank, likely because of the magnetic flux that threads the mutual inductance of the two surface mount components. Such crosstalk can be mitigated by separating tank-circuits that are close in resonant frequency or orienting inductors perpendicular to each other.

Finally, we evaluate the crosstalk between different layers of the device-PCB, with data shown in Fig. 3(d). Crosstalk is strongest for pairs of transmission lines that align or are  proximal with each other, despite being on separate PCB layers.  This is likely due to shared screening currents that flow in the common ground-plane. Additional coupling occurs at the end of the signal tracks where the ground layer has been removed to enable bonding to the chip.
\section{Crosstalk Mitigation}
Traditional sample-chip mounting methods, widely used in nano-electronics experiments, do not make use of the multi-layer circuit board techniques described here. The high density of interconnects needed for control and readout of mult-qubit devices however, leads to a crowding of high-frequency planar transmission lines at the PCB layer and a significant increase in crosstalk. In an effort to mitigate crosstalk when requiring large numbers of interconnects, we make use of continuous fencing-vias to tie together the ground planes above and below high-frequency transmission lines. This method essentially creates a quasi-coaxial grounding geometry surrounding the central conductor, as illustrated in Fig. 4(a) for the case of the coplanar transmission lines implemented on the high-frequency layer of the device-PCB. Using numerical simulation \cite{HFSSQ3D}, we evaluate the effect of these fencing-vias and find that they decrease crosstalk between closely aligned signal tracks by $\sim$ 25 dB, as shown in Fig. 4(b).

The residual crosstalk is largely due to the presence of screening currents flowing in the common ground between signal lines. Figures 4(c) and 4(d) show the result of a numerical simulation \cite{HFSSQ3D} for the electric and magnetic field density in a cross-section of the high-frequency layer of the device-PCB. The presence of fencing-vias can be seen to mitigate crosstalk by strongly suppressing the electric field between the transmission lines. A weaker suppression is seen for coupling produced by the magnetic field component.

\section{Discussion and Conclusion}
The use of multi-layer ground-planes and the fencing-via method analysed here represent an approach to mitigating crosstalk in PCB architectures that require a high density of wide bandwidth interconnects. We have not addressed the significant crosstalk and signal degradation that can occur on chip, but note that there are many approaches to suppressing crosstalk at the device layer \cite{Bronckers:2010vv}. As capacitance and mutual inductance between two conductors is proportion to their length, the contribution to crosstalk by on-chip structures can be small in comparison to the longer interconnect features required at the PCB layer. On-chip crosstalk can also be mimimized with the appropriate use of ground guards between signal-carrying transmission lines \cite{Su:1993vm,Bronckers:2010vv}. For qubit structures however, this is challenging as device operation requires significant capacitive coupling between surface electrodes and the quantum dot structures used to confine electron spin qubits. This direct cross-coupling then sets a lower bound for the level of indirect crosstalk tolerable at the PCB layer: it should be much less than the unavoidable coupling at the device layer, which is typically of the order of a few percent ($\sim$ -40 dB for GHz voltage pulses). We note that we have measured high-frequency crosstalk to be as high as $\sim$ -3 dB for a range of commercially available chip-mount packages commonly used in nanoelectronic and qubit experiments. 

In conclusion, we have described a coupled-PCB platform designed to enhance the operation and testing of spin qubit devices in the regime where a high density of interconnects are needed. The platform is well suited to frequent re-bonding of device chips with different geometries and performs well at cryogenic temperatures and in the presence of magnetic fields. Crosstalk is strongly suppressed below -40 dB  (1\% for voltage amplitudes) for all control and readout transmission lines by making use of a multi-layer device-PCB with alternating ground planes and fencing-vias. 

 \section{Acknowledgements}
 
We thank Xanthe Croot, Alice Mahoney, and Matthew Collins for technical assistance.  This research was supported by the IARPA/MQCO program and the U. S. Army Research Office under Contract No.  W911NF-11-1-0068 and the Australian Research Council Centre of Excellence Scheme (EQuS CE110001013).\\

* email: david.reilly@sydney.edu.au


\begin{thebibliography}{37}

\expandafter\ifx\csname natexlab\endcsname\relax\def\natexlab#1{#1}\fi
\expandafter\ifx\csname bibnamefont\endcsname\relax
  \def\bibnamefont#1{#1}\fi
\expandafter\ifx\csname bibfnamefont\endcsname\relax
  \def\bibfnamefont#1{#1}\fi
\expandafter\ifx\csname citenamefont\endcsname\relax
  \def\citenamefont#1{#1}\fi
\expandafter\ifx\csname url\endcsname\relax
  \def\url#1{\texttt{#1}}\fi
\expandafter\ifx\csname urlprefix\endcsname\relax\def\urlprefix{URL }\fi
\providecommand{\bibinfo}[2]{#2}
\providecommand{\eprint}[2][]{\url{#2}}

\bibitem[{\citenamefont{Loss and DiVincenzo}(1998)}]{Loss:1998via}
\bibinfo{author}{\bibfnamefont{D.}~\bibnamefont{Loss}} \bibnamefont{and}
  \bibinfo{author}{\bibfnamefont{D.}~\bibnamefont{DiVincenzo}},
  \bibinfo{journal}{Phys. Rev. A} \textbf{\bibinfo{volume}{57}},
  \bibinfo{pages}{120} (\bibinfo{year}{1998}).

\bibitem[{\citenamefont{Kane}(1998)}]{Kane:1998wha}
\bibinfo{author}{\bibfnamefont{B.~E.} \bibnamefont{Kane}},
  \bibinfo{journal}{Nature (London)} \textbf{\bibinfo{volume}{393}},
  \bibinfo{pages}{133} (\bibinfo{year}{1998}).

\bibitem[{\citenamefont{Hanson et~al.}(2007)\citenamefont{Hanson, Petta,
  Tarucha, and Vandersypen}}]{Hanson:2007eg}
\bibinfo{author}{\bibfnamefont{R.}~\bibnamefont{Hanson}},
  \bibinfo{author}{\bibfnamefont{J.~R.} \bibnamefont{Petta}},
  \bibinfo{author}{\bibfnamefont{S.}~\bibnamefont{Tarucha}}, \bibnamefont{and}
  \bibinfo{author}{\bibfnamefont{L.~M.~K.} \bibnamefont{Vandersypen}},
  \bibinfo{journal}{Rev. Mod. Phys.} \textbf{\bibinfo{volume}{79}},
  \bibinfo{pages}{1217} (\bibinfo{year}{2007}).

\bibitem[{\citenamefont{Petta et~al.}(2005)\citenamefont{Petta, Johnson,
  Taylor, Laird, Yacoby, Lukin, Marcus, Hanson, and Gossard}}]{Petta:2005kn}
\bibinfo{author}{\bibfnamefont{J.~R.} \bibnamefont{Petta}},
  \bibinfo{author}{\bibfnamefont{A.~C.} \bibnamefont{Johnson}},
  \bibinfo{author}{\bibfnamefont{J.~M.} \bibnamefont{Taylor}},
  \bibinfo{author}{\bibfnamefont{E.~A.} \bibnamefont{Laird}},
  \bibinfo{author}{\bibfnamefont{A.}~\bibnamefont{Yacoby}},
  \bibinfo{author}{\bibfnamefont{M.~D.} \bibnamefont{Lukin}},
  \bibinfo{author}{\bibfnamefont{C.~M.} \bibnamefont{Marcus}},
  \bibinfo{author}{\bibfnamefont{M.~P.} \bibnamefont{Hanson}},
  \bibnamefont{and} \bibinfo{author}{\bibfnamefont{A.~C.}
  \bibnamefont{Gossard}}, \bibinfo{journal}{Science}
  \textbf{\bibinfo{volume}{309}}, \bibinfo{pages}{2180} (\bibinfo{year}{2005}).

\bibitem[{\citenamefont{Koppens et~al.}(2006)\citenamefont{Koppens, Buizert,
  Tielrooij, Vink, Nowack, Meunier, Kouwenhoven, and
  Vandersypen}}]{Koppens:2006kz}
\bibinfo{author}{\bibfnamefont{F.~H.~L.} \bibnamefont{Koppens}},
  \bibinfo{author}{\bibfnamefont{C.}~\bibnamefont{Buizert}},
  \bibinfo{author}{\bibfnamefont{K.~J.} \bibnamefont{Tielrooij}},
  \bibinfo{author}{\bibfnamefont{I.~T.} \bibnamefont{Vink}},
  \bibinfo{author}{\bibfnamefont{K.~C.} \bibnamefont{Nowack}},
  \bibinfo{author}{\bibfnamefont{T.}~\bibnamefont{Meunier}},
  \bibinfo{author}{\bibfnamefont{L.~P.} \bibnamefont{Kouwenhoven}},
  \bibnamefont{and} \bibinfo{author}{\bibfnamefont{L.~M.~K.}
  \bibnamefont{Vandersypen}}, \bibinfo{journal}{Nature (London)}
  \textbf{\bibinfo{volume}{442}}, \bibinfo{pages}{3443} (\bibinfo{year}{2006}).

\bibitem[{\citenamefont{Nowack et~al.}(2007)\citenamefont{Nowack, Koppens,
  Nazarov, and Vandersypen}}]{Nowack:2007du}
\bibinfo{author}{\bibfnamefont{K.~C.} \bibnamefont{Nowack}},
  \bibinfo{author}{\bibfnamefont{F.~H.~L.} \bibnamefont{Koppens}},
  \bibinfo{author}{\bibfnamefont{Y.~V.} \bibnamefont{Nazarov}},
  \bibnamefont{and} \bibinfo{author}{\bibfnamefont{L.~M.~K.}
  \bibnamefont{Vandersypen}}, \bibinfo{journal}{Science}
  \textbf{\bibinfo{volume}{318}}, \bibinfo{pages}{1430} (\bibinfo{year}{2007}).

\bibitem[{\citenamefont{Foletti et~al.}(2009)\citenamefont{Foletti, Bluhm,
  Mahalu, Umansky, and Yacoby}}]{Foletti:2009hka}
\bibinfo{author}{\bibfnamefont{S.}~\bibnamefont{Foletti}},
  \bibinfo{author}{\bibfnamefont{H.}~\bibnamefont{Bluhm}},
  \bibinfo{author}{\bibfnamefont{D.}~\bibnamefont{Mahalu}},
  \bibinfo{author}{\bibfnamefont{V.}~\bibnamefont{Umansky}}, \bibnamefont{and}
  \bibinfo{author}{\bibfnamefont{A.}~\bibnamefont{Yacoby}},
  \bibinfo{journal}{Nature Physics} \textbf{\bibinfo{volume}{5}},
  \bibinfo{pages}{903} (\bibinfo{year}{2009}).

\bibitem[{\citenamefont{Elzerman et~al.}(2004)\citenamefont{Elzerman, Hanson,
  Van~Beveren, Witkamp, Vandersypen, and Kouwenhoven}}]{Elzerman:803014}
\bibinfo{author}{\bibfnamefont{J.~M.} \bibnamefont{Elzerman}},
  \bibinfo{author}{\bibfnamefont{R.}~\bibnamefont{Hanson}},
  \bibinfo{author}{\bibfnamefont{L.~H.~W.} \bibnamefont{Van~Beveren}},
  \bibinfo{author}{\bibfnamefont{B.}~\bibnamefont{Witkamp}},
  \bibinfo{author}{\bibfnamefont{L.~M.~K.} \bibnamefont{Vandersypen}},
  \bibnamefont{and} \bibinfo{author}{\bibfnamefont{L.~P.}
  \bibnamefont{Kouwenhoven}}, \bibinfo{journal}{Nature (London)}
  \textbf{\bibinfo{volume}{430}}, \bibinfo{pages}{431} (\bibinfo{year}{2004}).

\bibitem[{\citenamefont{Amasha et~al.}(2008)\citenamefont{Amasha, MacLean,
  Radu, Zumb{\"u}hl, Kastner, Hanson, and Gossard}}]{Amasha:2008ky}
\bibinfo{author}{\bibfnamefont{S.}~\bibnamefont{Amasha}},
  \bibinfo{author}{\bibfnamefont{K.}~\bibnamefont{MacLean}},
  \bibinfo{author}{\bibfnamefont{I.}~\bibnamefont{Radu}},
  \bibinfo{author}{\bibfnamefont{D.}~\bibnamefont{Zumb{\"u}hl}},
  \bibinfo{author}{\bibfnamefont{M.}~\bibnamefont{Kastner}},
  \bibinfo{author}{\bibfnamefont{M.}~\bibnamefont{Hanson}}, \bibnamefont{and}
  \bibinfo{author}{\bibfnamefont{A.}~\bibnamefont{Gossard}},
  \bibinfo{journal}{Phys. Rev. Lett.} \textbf{\bibinfo{volume}{100}},
  \bibinfo{pages}{046803} (\bibinfo{year}{2008}).

\bibitem[{\citenamefont{Barthel et~al.}(2009)\citenamefont{Barthel, Reilly,
  Marcus, Hanson, and Gossard}}]{Barthel:2009hx}
\bibinfo{author}{\bibfnamefont{C.}~\bibnamefont{Barthel}},
  \bibinfo{author}{\bibfnamefont{D.~J.} \bibnamefont{Reilly}},
  \bibinfo{author}{\bibfnamefont{C.~M.} \bibnamefont{Marcus}},
  \bibinfo{author}{\bibfnamefont{M.~P.} \bibnamefont{Hanson}},
  \bibnamefont{and} \bibinfo{author}{\bibfnamefont{A.~C.}
  \bibnamefont{Gossard}}, \bibinfo{journal}{Phys. Rev. Lett.}
  \textbf{\bibinfo{volume}{103}}, \bibinfo{pages}{160503}
  (\bibinfo{year}{2009}).

\bibitem[{\citenamefont{Morello et~al.}(2010)\citenamefont{Morello, Pla,
  Zwanenburg, Chan, Tan, Huebl, M{\"o}tt{\"o}nen, Nugroho, Yang, van Donkelaar
  et~al.}}]{Morello:2010ga}
\bibinfo{author}{\bibfnamefont{A.}~\bibnamefont{Morello}},
  \bibinfo{author}{\bibfnamefont{J.~J.} \bibnamefont{Pla}},
  \bibinfo{author}{\bibfnamefont{F.~A.} \bibnamefont{Zwanenburg}},
  \bibinfo{author}{\bibfnamefont{K.~W.} \bibnamefont{Chan}},
  \bibinfo{author}{\bibfnamefont{K.~Y.} \bibnamefont{Tan}},
  \bibinfo{author}{\bibfnamefont{H.}~\bibnamefont{Huebl}},
  \bibinfo{author}{\bibfnamefont{M.}~\bibnamefont{M{\"o}tt{\"o}nen}},
  \bibinfo{author}{\bibfnamefont{C.~D.} \bibnamefont{Nugroho}},
  \bibinfo{author}{\bibfnamefont{C.}~\bibnamefont{Yang}},
  \bibinfo{author}{\bibfnamefont{J.~A.} \bibnamefont{van Donkelaar}},
  \bibnamefont{{\it et~al.}}, \bibinfo{journal}{Nature (London)}
  \textbf{\bibinfo{volume}{467}}, \bibinfo{pages}{687} (\bibinfo{year}{2010}).

\bibitem[{\citenamefont{van Weperen et~al.}(2011)\citenamefont{van Weperen,
  Armstrong, Laird, Medford, Marcus, Hanson, and Gossard}}]{vanWeperen:2011fl}
\bibinfo{author}{\bibfnamefont{I.}~\bibnamefont{van Weperen}},
  \bibinfo{author}{\bibfnamefont{B.~D.} \bibnamefont{Armstrong}},
  \bibinfo{author}{\bibfnamefont{E.~A.} \bibnamefont{Laird}},
  \bibinfo{author}{\bibfnamefont{J.}~\bibnamefont{Medford}},
  \bibinfo{author}{\bibfnamefont{C.~M.} \bibnamefont{Marcus}},
  \bibinfo{author}{\bibfnamefont{M.~P.} \bibnamefont{Hanson}},
  \bibnamefont{and} \bibinfo{author}{\bibfnamefont{A.~C.}
  \bibnamefont{Gossard}}, \bibinfo{journal}{Phys. Rev. Lett.}
  \textbf{\bibinfo{volume}{107}}, \bibinfo{pages}{030506}
  (\bibinfo{year}{2011}).

\bibitem[{\citenamefont{Brunner et~al.}(2011)\citenamefont{Brunner, Shin,
  Obata, Pioro-Ladri{\`e}re, Kubo, Yoshida, Taniyama, Tokura, and
  Tarucha}}]{Brunner:2011gma}
\bibinfo{author}{\bibfnamefont{R.}~\bibnamefont{Brunner}},
  \bibinfo{author}{\bibfnamefont{Y.~S.} \bibnamefont{Shin}},
  \bibinfo{author}{\bibfnamefont{T.}~\bibnamefont{Obata}},
  \bibinfo{author}{\bibfnamefont{M.}~\bibnamefont{Pioro-Ladri{\`e}re}},
  \bibinfo{author}{\bibfnamefont{T.}~\bibnamefont{Kubo}},
  \bibinfo{author}{\bibfnamefont{K.}~\bibnamefont{Yoshida}},
  \bibinfo{author}{\bibfnamefont{T.}~\bibnamefont{Taniyama}},
  \bibinfo{author}{\bibfnamefont{Y.}~\bibnamefont{Tokura}}, \bibnamefont{and}
  \bibinfo{author}{\bibfnamefont{S.}~\bibnamefont{Tarucha}},
  \bibinfo{journal}{Phys. Rev. Lett.} \textbf{\bibinfo{volume}{107}},
  \bibinfo{pages}{146801} (\bibinfo{year}{2011}).

\bibitem[{\citenamefont{Nowack et~al.}(2011)\citenamefont{Nowack, Shafiei,
  Laforest, Prawiroatmodjo, Schreiber, Reichl, Wegscheider, and
  Vandersypen}}]{Nowack:2011}
\bibinfo{author}{\bibfnamefont{K.~C.} \bibnamefont{Nowack}},
  \bibinfo{author}{\bibfnamefont{M.}~\bibnamefont{Shafiei}},
  \bibinfo{author}{\bibfnamefont{M.}~\bibnamefont{Laforest}},
  \bibinfo{author}{\bibfnamefont{G.~E. D.~K.} \bibnamefont{Prawiroatmodjo}},
  \bibinfo{author}{\bibfnamefont{L.~R.} \bibnamefont{Schreiber}},
  \bibinfo{author}{\bibfnamefont{C.}~\bibnamefont{Reichl}},
  \bibinfo{author}{\bibfnamefont{W.}~\bibnamefont{Wegscheider}},
  \bibnamefont{and} \bibinfo{author}{\bibnamefont{Vandersypen}},
  \bibinfo{journal}{Science} \textbf{\bibinfo{volume}{333}},
  \bibinfo{pages}{1269} (\bibinfo{year}{2011}).

\bibitem[{\citenamefont{Su et~al.}(1993)\citenamefont{Su, Loinaz, and
  Masui}}]{Su:1993vm}
\bibinfo{author}{\bibfnamefont{D.}~\bibnamefont{Su}},
  \bibinfo{author}{\bibfnamefont{M.}~\bibnamefont{Loinaz}}, \bibnamefont{and}
  \bibinfo{author}{\bibfnamefont{S.}~\bibnamefont{Masui}},
  \bibinfo{journal}{IEEE J. Solid-state Circuits}
  \textbf{\bibinfo{volume}{28}}, \bibinfo{pages}{420} (\bibinfo{year}{1993}).

\bibitem[{\citenamefont{Preskill}(1998)}]{Preskill:1998vu}
\bibinfo{author}{\bibfnamefont{J.}~\bibnamefont{Preskill}},
  \bibinfo{journal}{Proc. R. Soc. Lond. A} \textbf{\bibinfo{volume}{454}},
  \bibinfo{pages}{385} (\bibinfo{year}{1998}).

\bibitem[{\citenamefont{Gustavsson et~al.}(2008)\citenamefont{Gustavsson,
  Shorubalko, Leturcq, Ihn, Ensslin, and Sch{\"o}n}}]{Gustavsson:2008gb}
\bibinfo{author}{\bibfnamefont{S.}~\bibnamefont{Gustavsson}},
  \bibinfo{author}{\bibfnamefont{I.}~\bibnamefont{Shorubalko}},
  \bibinfo{author}{\bibfnamefont{R.}~\bibnamefont{Leturcq}},
  \bibinfo{author}{\bibfnamefont{T.}~\bibnamefont{Ihn}},
  \bibinfo{author}{\bibfnamefont{K.}~\bibnamefont{Ensslin}}, \bibnamefont{and}
  \bibinfo{author}{\bibfnamefont{S.}~\bibnamefont{Sch{\"o}n}},
  \bibinfo{journal}{Phys. Rev. B} \textbf{\bibinfo{volume}{78}},
  \bibinfo{pages}{035324} (\bibinfo{year}{2008}).

\bibitem[{\citenamefont{Kouwenhoven et~al.}(1994)\citenamefont{Kouwenhoven,
  Jauhar, and Orenstein}}]{Kouwenhoven:1994ww}
\bibinfo{author}{\bibfnamefont{L.~P.} \bibnamefont{Kouwenhoven}},
  \bibinfo{author}{\bibfnamefont{S.}~\bibnamefont{Jauhar}}, \bibnamefont{and}
  \bibinfo{author}{\bibfnamefont{J.}~\bibnamefont{Orenstein}},
  \bibinfo{journal}{Phys. Rev. Lett.} \textbf{\bibinfo{volume}{73}},
  \bibinfo{pages}{3443} (\bibinfo{year}{1994}).

\bibitem[{\citenamefont{Switkes et~al.}(1999)\citenamefont{Switkes, Marcus, and
  Campman}}]{Switkes:1999uva}
\bibinfo{author}{\bibfnamefont{M.}~\bibnamefont{Switkes}},
  \bibinfo{author}{\bibfnamefont{C.~M.} \bibnamefont{Marcus}},
  \bibnamefont{and} \bibinfo{author}{\bibfnamefont{K.}~\bibnamefont{Campman}},
  \bibinfo{journal}{Science} \textbf{\bibinfo{volume}{283}},
  \bibinfo{pages}{1905} (\bibinfo{year}{1999}).

\bibitem[{\citenamefont{Khodjasteh and Lidar}(2007)}]{Khodjasteh:2007bu}
\bibinfo{author}{\bibfnamefont{K.}~\bibnamefont{Khodjasteh}} \bibnamefont{and}
  \bibinfo{author}{\bibfnamefont{D.}~\bibnamefont{Lidar}},
  \bibinfo{journal}{Phys. Rev. A} \textbf{\bibinfo{volume}{75}},
  \bibinfo{pages}{062310} (\bibinfo{year}{2007}).

\bibitem[{Inn()}]{coax}
\bibinfo{note}{Micro-Coax Corp. UT85 Inner-Outer SC-219/50-SS-SS and 0.86 mm CuNi Silver-plated CuNi SC-086/50-SCN-CN.}

\bibitem[{Lei()}]{Leiden}
  \bibinfo{note}{Leiden Cryogenics. Dry dilution refrigerator type
  CF-450.}

\bibitem[{\citenamefont{Emerson Cuming Eccosorb MF~117}()}]{Eccosorb}
\bibinfo{author}{\bibfnamefont{D. F.}~\bibnamefont{Santavicca}} \bibnamefont{and}
 \bibinfo{author}{\bibfnamefont{D. E.}~\bibnamefont{Prober}},
 \bibinfo{journal}{Meas. Sci. Technol.} \textbf{\bibinfo{volume}{19}},
 \bibinfo{pages}{087001} (\bibinfo{year}{2008}).
\bibinfo{author}{\bibnamefont{Emerson \& Cuming Corp. Eccosorb
  MF~117}}.

\bibitem[{\citenamefont{{\textnormal{Glenair Inc. Series 89 Nanominiature
  Connectors 890-013}}}()}]{nano}
\bibinfo{author}{\bibnamefont{{\textnormal{Glenair Corp. Series 89 Nanominiature
  Connectors 890-013}}}}.

\bibitem[{smp()}]{smp}
SMP Connectors, 18 S 142- 40M L5 and SMP bullets 18 K 101- K00 L5, Rosenberger Corp.


\bibitem[{\citenamefont{Tischler et~al.}(2003)\citenamefont{Tischler, Rudolph,
  and Kilk}}]{Tischler:2003tp}
\bibinfo{author}{\bibfnamefont{T.}~\bibnamefont{Tischler}},
  \bibinfo{author}{\bibfnamefont{M.}~\bibnamefont{Rudolph}}, \bibnamefont{and}
  \bibinfo{author}{\bibfnamefont{A.}~\bibnamefont{Kilk}},
  \bibinfo{journal}{IEEE MTT-S International Microwave Symp. Digest}
  \textbf{\bibinfo{volume}{2}}, \bibinfo{pages}{1147} (\bibinfo{year}{2003}).

\bibitem[{\citenamefont{Yuasa and Nishino}(2004)}]{Yuasa:2004uf}
\bibinfo{author}{\bibfnamefont{T.}~\bibnamefont{Yuasa}} \bibnamefont{and}
  \bibinfo{author}{\bibfnamefont{T.}~\bibnamefont{Nishino}},
  \bibinfo{journal}{IEEE MTT-S International Microwave Symp. Digest}
  \textbf{\bibinfo{volume}{2}}, \bibinfo{pages}{641} (\bibinfo{year}{2004}).

\bibitem[{\citenamefont{Imai et~al.}(2009)\citenamefont{Imai, Sugimura, and
  Kawasaki}}]{Imai:2009tn}
\bibinfo{author}{\bibfnamefont{H.}~\bibnamefont{Imai}},
  \bibinfo{author}{\bibfnamefont{M.}~\bibnamefont{Sugimura}}, \bibnamefont{and}
  \bibinfo{author}{\bibfnamefont{M.}~\bibnamefont{Kawasaki}},
  \bibinfo{journal}{IEEE Trans. Components and Packaging Tech.}
  \textbf{\bibinfo{volume}{32}}, \bibinfo{pages}{415} (\bibinfo{year}{2009}).

\bibitem[{min()}]{mini}
 Minicircuits Corp. Filters LFCN-80+ 

\bibitem[{\citenamefont{Ponchak and Chen}(1998)}]{Ponchak:1998tm}
\bibinfo{author}{\bibfnamefont{G.}~\bibnamefont{Ponchak}} \bibnamefont{and}
  \bibinfo{author}{\bibfnamefont{D.}~\bibnamefont{Chen}},
  \bibinfo{journal}{IEEE MTT-S International Microwave Symp. Digest}
  \textbf{\bibinfo{volume}{3}}, \bibinfo{pages}{1831} (\bibinfo{year}{1998}).

\bibitem[{\citenamefont{Wu and Kermani}(2003)}]{Wu:2003wa}
\bibinfo{author}{\bibfnamefont{X.}~\bibnamefont{Wu}} \bibnamefont{and}
  \bibinfo{author}{\bibfnamefont{M.}~\bibnamefont{Kermani}},
  \bibinfo{journal}{IEEE Electrical Performance of Electronic Packaging} p.
  \bibinfo{pages}{207} (\bibinfo{year}{2003}).

\bibitem[{\citenamefont{textnormal Southwest~Microwave}()}]{esr}
\bibinfo{author}{\bibnamefont{Southwest Microwave Corp. Connectors: 2.40 mm 50 GHz 1492-02A-5.}}

\bibitem[{\citenamefont{Reilly et~al.}(2007)\citenamefont{Reilly, Marcus,
  Hanson, and Gossard}}]{Reilly:2007ig}
\bibinfo{author}{\bibfnamefont{D.~J.} \bibnamefont{Reilly}},
  \bibinfo{author}{\bibfnamefont{C.~M.} \bibnamefont{Marcus}},
  \bibinfo{author}{\bibfnamefont{M.~P.} \bibnamefont{Hanson}},
  \bibnamefont{and} \bibinfo{author}{\bibfnamefont{A.~C.}
  \bibnamefont{Gossard}}, \bibinfo{journal}{App. Phys. Lett.}
  \textbf{\bibinfo{volume}{91}}, \bibinfo{pages}{162101}
  (\bibinfo{year}{2007}).

\bibitem[{\citenamefont{Agilent~PNA5230C}()}]{PNA}
\bibinfo{author}{\bibnamefont{Agilent Corp. ~PNA5230C}}.

\bibitem[{Lak()}]{Lakeshore}
LakeShore Cryotronics Corp. Model CRX-4K.

\bibitem[{\citenamefont{numerical~simulation software}()}]{HFSSQ3D}
\bibinfo{author}{\bibnamefont{HFSS Ansoft Corp. and Q3D extractor.}} 

\bibitem[{\citenamefont{Bronckers et~al.}(2010)\citenamefont{Bronckers, Van
  Der~Plas, and Vandersteen}}]{Bronckers:2010vv}
\bibinfo{author}{\bibfnamefont{S.}~\bibnamefont{Bronckers}},
  \bibinfo{author}{\bibfnamefont{G.}~\bibnamefont{Van Der~Plas}},
  \bibnamefont{and}
  \bibinfo{author}{\bibfnamefont{G.}~\bibnamefont{Vandersteen}},
  \emph{\bibinfo{title}{{Substrate noise coupling in Analog/RF circuits. Artech
  House Publishers}}} (\bibinfo{year}{2010}).

\end{thebibliography}
\end{document}